# Rich magnetic phase diagram in the Kagome-staircase compound $Mn_3V_2O_8$

E. Morosan,[1] J. Fleitman,[1] T. Klimczuk,[2,3] and R. J. Cava[1]

[1]Department of Chemistry, Princeton University, Princeton NJ 08544

[2]Division of Thermal Physics, Los Alamos National Laboratories, Los Alamos NM 87545

[3]Faculty of Applied Physics and Mathematics, Gdansk University of Technology, Narutowicza 11/12, 80-952 Gdansk, Poland

$Mn_3V_2O_8$ is a magnetic system in which $S$ = 5/2 $Mn^{2+}$ is found in the kagomé staircase lattice. Here we report the magnetic phase diagram for temperatures above 2 K and applied magnetic fields below 9 T, characterized by measurements of the magnetization and specific heat with field along the three unique lattice directions. At low applied magnetic fields, the system first orders magnetically below $T_{m1} \approx 21$ K, and then shows a second magnetic phase transition at $T_{m2} \approx 15$ K. In addition, a phase transition that is apparent in specific heat but not seen in magnetization is found for all three applied field orientations, converging towards $T_{m2}$ as $H \rightarrow 0$. The magnetic behavior is highly anisotropic, with critical fields for magnetic phase boundaries much higher when the field is applied perpendicular to the Kagomé staircase plane than when applied in-plane. The field-temperature ($H$ - $T$) phase diagrams are quite rich, with 7 distinct phases observed.

Geometrically frustrated magnetic materials have recently emerged as the focus of intense study. Among these, compounds based on the kagomé net, a regular planar lattice made from corner sharing of equilateral triangles, are of particular interest due to the very high degeneracy of energetically equivalent magnetic ground states. Breaking the ideal triangular symmetry of the kagomé net typically favors one particular magnetically ordered state above others. For the particular case of the kagomé-staircase geometry, however, in which the symmetry breaking occurs via buckling of the kagomé plane (see inset to Fig. 1), an exquisitely close competition between different magnetically ordered states has been observed, resulting in complex temperature-applied field magnetic phase diagrams. The kagomé-staircase lattice is observed in the transition metal vanadates $T_3V_2O_8$ (with T = Co, Ni, Cu, and Zn) [1-11], and the Ni [2-4,7-8,11] and Co [1-3,5-6] variants have been widely studied. Simultaneous long-range ferroelectric and magnetic order have been found in $Ni_3V_2O_8$ [11], allowing its classification as a multiferroic compound.

Orthorhombic symmetry $Mn_3V_2O_8$ (MVO) is isostructural with $Co_3V_2O_8$ (CVO) and $Ni_3V_2O_8$ (NVO), but its physical properties have been only marginally studied [12]. The $t_{2g}^3e_g^2$, isotropic spin, $S = 5/2$ $L = 0$ configuration of $Mn^{2+}$ in MVO presents an interesting contrast to the $t_{2g}^5e_g^2$ $S = 3/2$ $Co^{2+}$ and $t_{2g}^6e_g^2$ $S = 1$ $Ni^{2+}$ cases of CVO and NVO. Here we report the observation of rich anisotropic magnetic field-temperature ($H$ - $T$) phase diagrams for MVO, as determined from magnetization and specific heat measurements on single crystals. Two distinct magnetic phase transitions, at 21 K and 15 K, are observed for fields applied in all three principal crystallographic directions. A phase transition that is apparent in specific heat but not seen in magnetization is found for all three applied field orientations, converging towards the 15 K transition as $H \rightarrow 0$. The magnetic behavior is highly anisotropic, with critical fields for magnetic phase boundaries much higher when the magnetic field is applied perpendicular to the kagomé-staircase plane (*H* parallel to the *b* crystallographic axis) then when applied in-plane (*H* parallel to the *a* and *c* crystallographic axes). The field-temperature (*H* - *T*) phase diagrams are quite rich, with 7 distinct magnetic/structural phases observed. The magnetic phase diagrams are distinctly different from what is observed for CVO and NVO.

Single crystals of MVO were grown out of a $MoO_3/V_2O_5/MnO$ flux as previously described [12]. The starting oxides (MnO 99% Aldrich, $V_2O_5$ 99.6% Alfa Aesar, $MoO_3$ 99.95% Alfa Aesar) were packed in an alumina crucible, which was then heated in a vertical tube furnace under flowing Argon gas. Sacrificial MnO powder was placed in an alumina crucible above the $MoO_3/V_2O_5/MnO$ flux to create an oxygen partial pressure that would neither oxidize the $Mn^{2+}$ nor reduce the $V^{5+}$[ref 12]. The vertical furnace was heated to 1200 $^{o}$C at 200 $^{o}$C/hr, held at 1200 $^{o}$C for 1 hour, cooled to 900 $^{o}$C at 5 $^{o}$C /hr, then cooled to room temperature at 300 $^{o}$C/hr. After the heat treatment red-brown platelet crystals were extracted from the flux using a bath of 1 part glacial acetic acid (Fisher) and 3 parts deionized water. The crystals were found to be single phase by single crystal and powder X-ray diffraction, with the orthorhombic *Cmca* structure and lattice parameters $a$ = 6.2672(3) Å, $b$ = 11.7377(8) Å and $c$ = 8.5044(5) Å. Field- and temperature-magnetization measurements were performed in a Quantum Design Physical Properties Measurement System (PPMS). The specific heat data were also collected in a PPMS, using a relaxation technique with fitting of the whole calorimeter (sample with sample platform and puck).

The $H$ = 0.5 T inverse magnetization data for MVO (Fig.1) indicates the presence of long range magnetic ordering below $T_{m1} \approx 21$ K. Previous low-field magnetization data [12] suggest the existence of an additional magnetic phase transition near 40 K, but the feature observed is most likely due to the presence of an $Mn_2V_2O_7$ impurity phase. A high-temperature fit of the susceptibility (dotted line, Fig.1) to the Curie-Weiss law $\chi = \chi_0 + C/(T-\theta_W)$ yields an effective moment $\mu_{eff} = 5.94\ \mu_B$, in excellent agreement with the theoretical value $\mu_{eff} = 5.92\ \mu_B$ expected for high-spin $S = 5/2$ $Mn^{2+}$. The Weiss temperature $\theta_W = -320$ K indicates the dominance of antiferromagnetic exchange interactions. Given the kagomé staircase magnetic lattice (inset in Fig. 1), it is not surprising that $|\theta_W / T_{m1}| \approx 15$, characteristic of a strongly frustrated antiferromagnetic spin system. Deviations from Curie-Weiss behavior, typical of magnetically frustrated materials, are observed to begin on cooling at approximately 70 K.

The easy magnetization axis lies close to the kagomé-staircase *ac*-plane, where the magnetization is largest (Fig.1). Upon further inspection of the behavior of the magnetization in different applied fields, two magnetic phase transitions can be identified in the M(T) data for all field orientations. Fig. 2a and c illustrate the field dependence of these transitions for H || *a* and H || *c* respectively. The competition between the antiferromagnetic spin coupling and the anisotropy associated with the kagomé staircase structure precludes the system from attaining a zero net magnetization ground state. This is suggested by the rapidly increasing magnetization as the system enters the high temperature, low field state (HT1) upon cooling below $T_{m1} \approx 21$ K. A net ferromagnetic component can probably be associated with the HT1 phase. Subsequent cooling of the sample gives rise to a sharp cusp followed by a local minimum around $T_{m2} = 15$ K, where a second magnetic phase transition, from HT1 to a low-T, low-H state (LT1) occurs. Increasing magnetic field (Fig.2a,c) has almost no effect on the long-range magnetic ordering temperature $T_{m1}$, but it broadens the cusp and slowly drives the second transition down in temperature. Above $H = 0.04$ T the H || *a* low temperature magnetization plateaus at a finite value, which strongly suggests a canted spin configuration even for the LT1 state, with a smaller ferromagnetic component along *a* than in the HT1 state. Very similar behavior occurs for the other in-plane orientation H || *c* (Fig.2c), with the two distinct transitions persisting up to slightly higher field $H$ = 0.1 T. The insets in Fig. 2a,c represent examples of how the critical temperatures for the magnetic phase transitions at constant field are determined: as shown, the vertical arrows mark $T_{m1}$ and $T_{m2}$, $H$ = 0.01 T, and correspond to local minima in the temperature-derivative of magnetization d$M$/d$T$. The two low-field phases that are observed in the magnetically ordered state are possibly a result of the ordering of the spins on one or both of the distinct $Mn^{2+}$ ions (inset Fig. 1) in the *ac*-plane, similar to the transitions encountered in NVO [4]. Fewer magnetic phases are distinguishable at low fields in MVO, however, than are seen in either NVO or CVO.

A more complex scenario is revealed in MVO at finite fields. Fig. 2b shows a selection of the H || *b* $M$(H) isotherms (full symbols), with the $T$ = 2 K field-derivative d$M$/d$H$ (open diamonds) as an example of how the critical field values were determined. At $T$ = 2 K, the magnetization is low and linear with field for $H$ < 2 T, which corresponds to the LT1 phase. This behavior is consistent with the antiferromagnetic spins slowly rotating from the easy axis in the *ac* plane, closer to the direction of the applied field H || *b*. A sharp step in magnetization around $H_{c1}$ = 2.1 T marks the transition from LT1 to LT2, possibly due to a spin-flop transition on one or both of the $Mn^{2+}$ sites. Although the magnetization increases linearly with field above this transition, as expected for the spin-reorientation subsequent to a spin-flop, another transition occurs just below 3 T, where $M$(H) changes slope (full diamonds, Fig.2b) and the system enters the state LT3. The spin-flop transition yields a sharp peak in d$M$/d$H$ (open diamonds, Fig.2b); the higher critical field value $H_{c2}$ is determined using an on-set criterion for d$M$/d$H$. Both transitions are marked by small vertical arrows in Fig.2b.

As the temperature is raised, the initial slope of the *M*(H) curves increases in the magnetically ordered phase (Fig.2b) such that the magnetization jump at the spin-flop transition becomes indistinguishable. The two magnetic phase transitions move slightly down with field, and are hard to identify in the magnetization measurements above 16 K. Specific heat measurements complement the magnetization data, by confirming the magnetic phase lines, but also by revealing another phase transition that was not visible in the *M*(T,H) data. A selection of the specific heat curves, plotted as $C_p/T$ *vs.* $T$, is shown in Fig. 3a, for H || *b* and applied fields up to 9 T. For $H = 0$ (full squares) a sharp peak associated with long range magnetic ordering is seen around $T_{m1}$ = 21 K, with a second peak at the lower phase transition temperature $T_{m2}$. After subtracting the lattice contribution to the specific heat as measured for the non-magnetic analogous compound $Zn_3V_2O_8$ [2] (solid line, Fig.3b) one can estimate the magnetic specific heat $C_m$ for MVO (open symbols, Fig.3b). The temperature dependence of the magnetic entropy can then be calculated and is shown in the inset in Fig.3b for $H$ = 0 (open circles): only about 50% of the R *ln*6 entropy expected for a $S$ = 5/2 state is accounted for between 2 and 40 K. This could be an indication that additional phase transitions may exist below 2 K. Another possible explanation, given the observed departure from Curie-Weiss behavior below ~ 70 K (Fig.1), is that more entropy is associated with short range order below 70 K. No additional entropy is recovered with the application of magnetic field, as the $H$ = 9 T temperature-dependent entropy (crosses, inset Fig.3b) differs only slightly from the $H$ = 0 data. However, as the field is turned on, very different behavior is observed for the two peaks in $C_p$ (Fig.3a): the one just below 16 K is affected little in temperature by the increasing magnetic field, but the higher-temperature one moves down in field. Concurrently, a third, broader peak emerges above ~ 1.5 T and is driven higher in temperature with increasing field. It is likely that both phase transitions exist at finite fields even for $H < 1.5$ T, and converge at $T_{m1}$ for H → 0, but their proximity in temperature makes it impossible to discern two separate peaks. For $H > H^b_{c1}$ the lower temperature peak is not associated with any phase transition observed in the magnetization data. Given its invisibility in the magnetization, and the relative insensitivity of the transition temperature to applied field, we speculate that this transition may have a structural component, though the fact that the amount of entropy in the transition is suppressed by the field indicates that there must be a magnetic component as well.

Based on our extensive magnetization and specific heat measurements, we present the H – T magnetic phase diagrams for MVO, for magnetic fields applied along the unique structural directions, in Figs. 4 and 5. As the temperature is lowered in zero field, MVO orders magnetically at $T_{m1} = (20.7 \pm 0.2)$ K, entering first a high temperature phase (HT1) and then a low temperature phase (LT1) at $T_{m2} = (15.2 \pm 0.5)$ K. The response of the system to applied magnetic field is highly anisotropic. For H || $a$ (Fig. 4a), in finite field, two distinct phase boundaries emerge at $T_{m2}$: one represents the lower temperature magnetic phase transition, which moves down in temperature as H increases, and the second is an almost vertical line, which is only visible in the specific heat data. The intermediate temperature phase delineated by these two phase boundaries is LT4, which extends in field up to about 0.04 T. An almost horizontal phase line cuts across the phase diagram at $H^a_{c1} \approx 0.04$ T. It separates the low field, low temperature (LT1) and a high temperature (HT1) phases from two different states (LT3 and HT3) at higher fields.

For the other field orientation close to the plane (H || $c$, Fig. 4b), the low field phase diagram is similar to that for H || $a$, with the HT1, LT4 and LT1 phases extending up in field up to a much higher critical value $H^c_{c1} = 0.3$ T. In the T → 0 limit, a second magnetic phase

transition occurs at $H^{c}_{c2}$ = 2.6 T, and the critical field value is slowly reduced with temperature. The two almost horizontal phase lines at $H^{c}_{c1}$ and $H^{c}_{c2}$ separate a low temperature (LT2) and a high temperature (HT2) phase at intermediate field values from the high field states LT3 and HT3.

When field is applied perpendicular to the kagomé planes (Fig. 5) the phase diagram is analogous to the in-plane ones. The most noticeable difference is that the critical field values are much higher: $H^{b}_{c1}$ = 2.2 T and $H^{b}_{c2}$ = 3.0 T respectively for T → 0. This is expected given the observed anisotropy, which constrains the magnetic moments to lie closer to the *ac*-plane: stronger fields are needed to pull the moments towards the "hard" axis *b*. In addition, the LT4 phase is missing, and the phase line that starts at $H^{b}_{c1}$ at T → 0 converges at $T_{m2}$ in the H = 0 limit. As a consequence, the HT2 phase merges with HT1 just below the magnetic ordering at $T_{m1}$.

The temperature-field magnetic phase diagram for $Mn_3V_2O_8$ is quite different from those seen in $Ni_3V_2O8$ and $Co_3V_2O_8$. In all three compounds, the competition between the crystalline anisotropy and the antiferromagnetic interactions in the kagomé staircase structure gives rise to strong geometric frustration. In NVO and CVO, differences in the magnetically ordered states have been found to involve differences in the ordering of the moments on the two kinds of magnetic ion sites, the so-called spine and crosstie sites. The same will no doubt prove true for MVO, with the present measurements revealing that the magnetic moments on the two distinct $Mn^{2+}$ sites lie close to the *ac*-plane when in the H = 0 magnetically ordered states. For magnetic fields applied in-plane, the magnetic states in MVO are much more sensitive to applied field than they are in NVO and CVO, with fig. 4 showing for example that the LT1 and HT1 phases disappear in applied fields in the *a* direction as low as 0.04 T. The complexity of the anisotropic $H - T$ phase diagrams in MVO appears to be derived from competition between nearly balanced magnetic interactions, leading to canted spin configurations or field-induced spin-flop transitions. An integration of the entropy observed under the $H = 0$ phase transitions between 2 and 40 K does not yield the expected R*ln*6 for $Mn^{2+}$, suggesting that there may be more magnetic phase transitions below 2 K, or that additional entropy is associated with short-range order below 70 K. Detailed neutron scattering measurements are desirable in order to elucidate the nature of the different states observed in MVO, and also to clarify whether the almost field independent phase boundary at $T_{m2}$ is associated with a structural phase transition. Investigation of possible multiferroic phases will also be of considerable interest.

**Acknowledgements**

This research was supported by the US Department of Energy, Division of Basic Energy Sciences, grant DE-FG02- 98-ER45706. We thank G. Lawes for providing the specific heat data for $Zn_3V_2O_8$.

**Figure Captions**

Fig 1. Anisotropic inverse susceptibility data for *H* = 0.5 T (symbols) and linear fit of the high-temperature data (dotted line). Insert: kagomé staircase structure of the $Mn^{2+}$ array in $Mn_3V_2O_8$. Crystallographic axes are shown. Spine sites are shown in purple and crosstie sites are shown in pink.

Fig. 2. (a) H || *a* *M*(T) data for *H* = 0.015 T, 0.025 T, 0.04 T, 0.06 T, 0.08 T, 0.1 T and 5.0 T. Inset: the *H* = 0.01 T d*M*/d*T* curve; small vertical arrows indicate the position of the minima in the derivative, from which the critical temperature values are determined. (b) H || *b* *M*(H) isotherms for *T* = 2 K, 15 K, 18 K, 20 K and 30 K (full symbols, left axis); the *T* = 2 K d*M*/d*H* curve (open symbols, right axis) illustrates how the critical field values $H_{c1}$ and $H_{c2}$, marked by vertical arrows, are determined. (c) H || *c* *M*(T) data for *H* = 0.02 T, 0.05 T, 0.5 T, 1.0 T, 2.0 T, 3.0 T and 5.0 T. Inset: the *H* = 0.01 T d*M*/d*T* curve; small vertical arrows indicate the position of the minima in the derivative, from which the critical temperature values are determined.

Fig 3. (a) H || *b* $C_p/T$ *vs.* *T* data for *H* = 0, 1.5 T, 2.0 T, 6.0 T and 9.0 T. (b) $C_p/T$ data for MVO (full symbols) and $Zn_3V_2O_8$ (solid line) (right axis) used to determine the magnetic specific heat $C_m$ of MVO (open symbols, left axis) plotted as $C_m/T$. Inset: the temperature-dependence of the magnetic entropy $S_m$ for *H* = 0 and 9 T.

Fig 4. (a) H || *a* and (b) H || *c* *H* – *T* phase diagrams: points are determined from *M*(T) data (orange symbols), *M*(H) data (wine symbols) or $C_p(T)|_H$ data (open symbols). The solid lines are guides connecting the points determined experimentally; extrapolations of these phase

boundaries in regions where measurements were missing or critical *H* and *T* values were difficult to determine are represented by dotted lines.

Fig 5. H || *b* $H-T$ phase diagrams: points are determined from *M*(T) data (orange symbols), *M*(H) data (wine symbols) or $C_p(T)|_H$ data (open symbols). The solid lines are guides connecting the points determined experimentally; extrapolations of these phase boundaries in regions where measurements were missing or critical *H* and *T* values were difficult to determine are represented by dotted lines.

Fig.1.

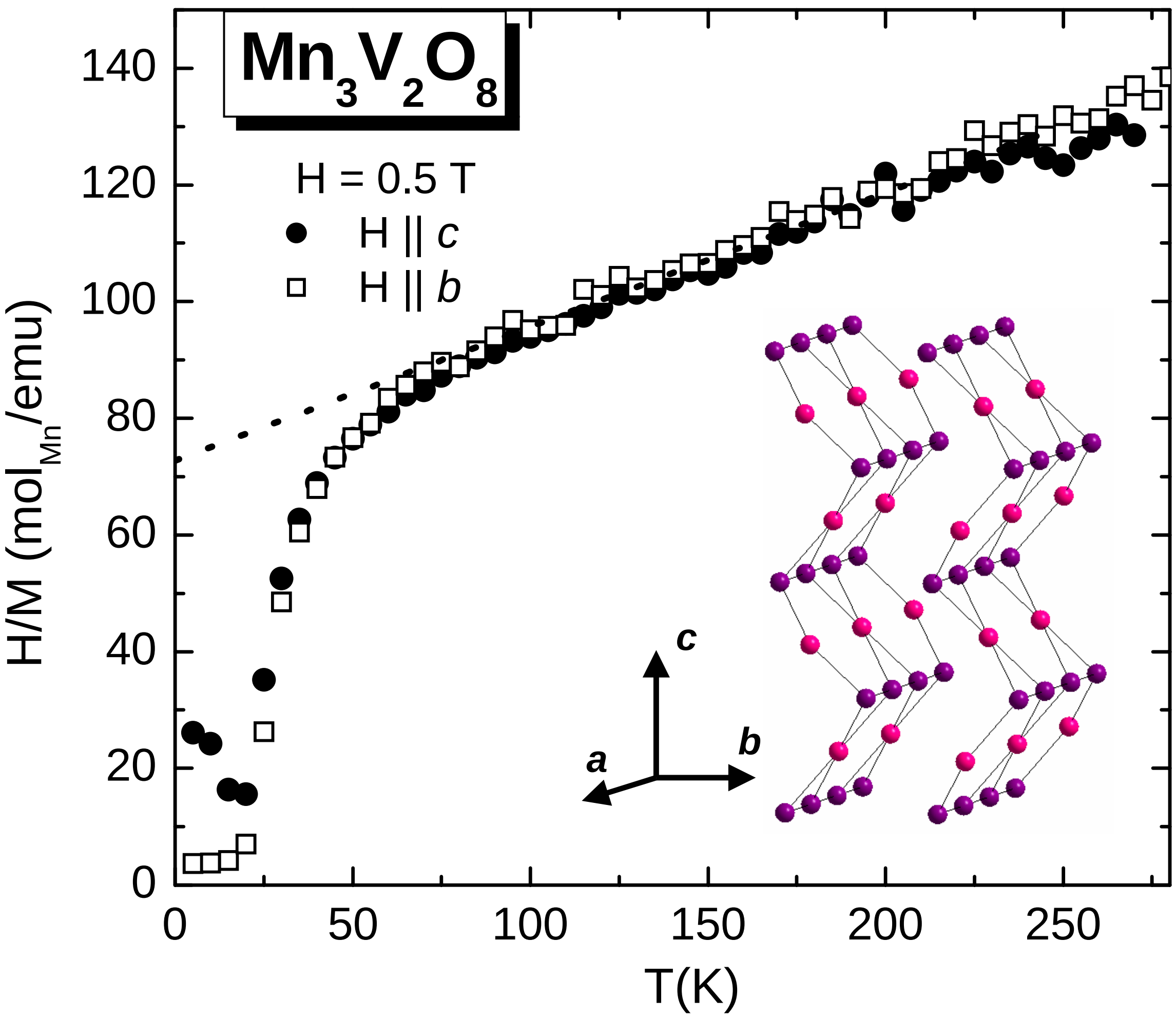

Fig. 2.

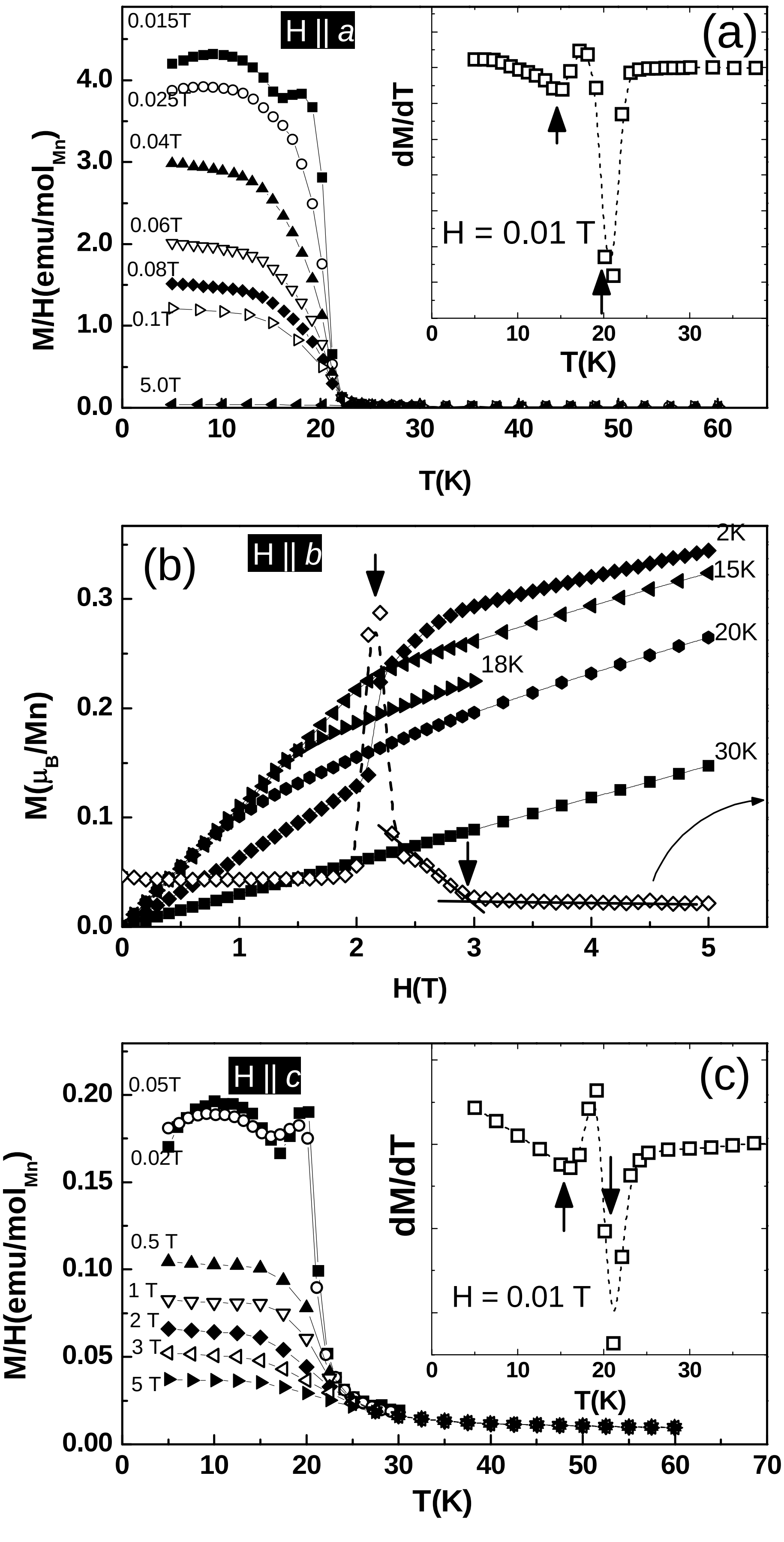

Fig. 3.

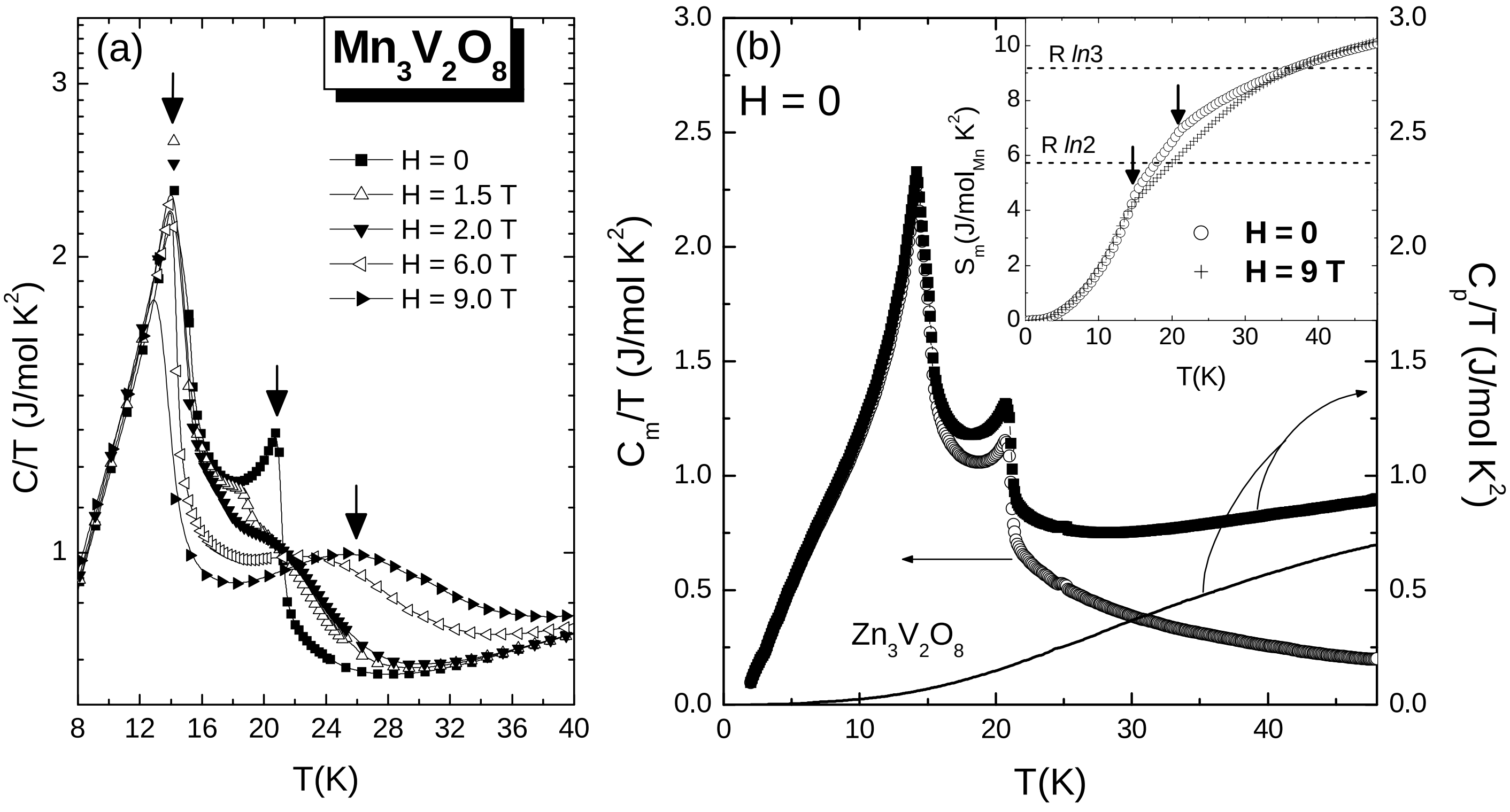

(a)
Mn3V2O8
H = 0
H = 1.5 T
H = 2.0 T
H = 6.0 T
H = 9.0 T
C/T (J/mol K2)
T(K)
(b)
H = 0
Cm/T (J/mol K2)
Cp/T (J/mol K2)
Zn3V2O8
T(K)
R ln3
R ln2
Sm(J/molMn K2)
H = 0
H = 9 T
T(K)

Fig.4.

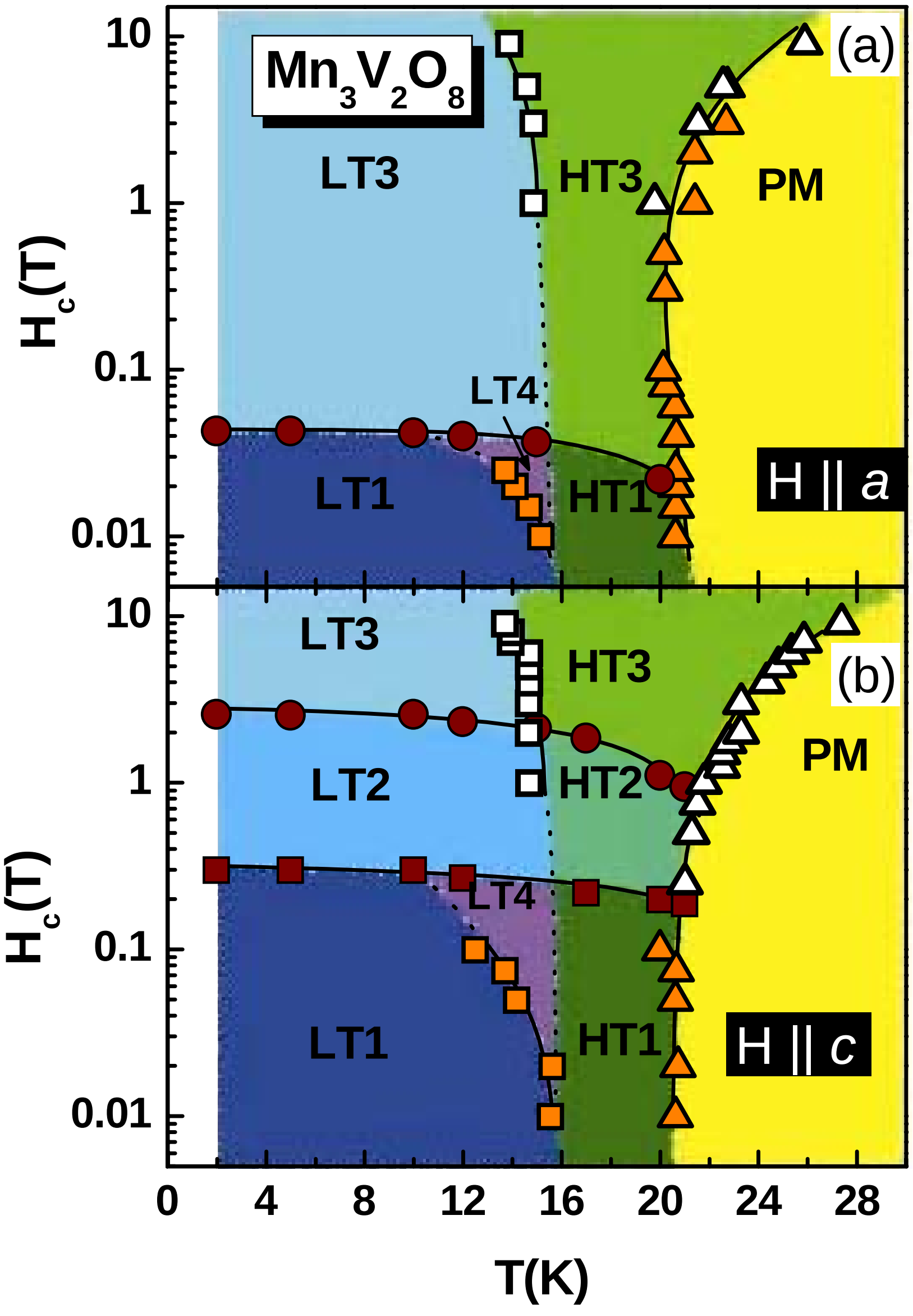

Fig.5.

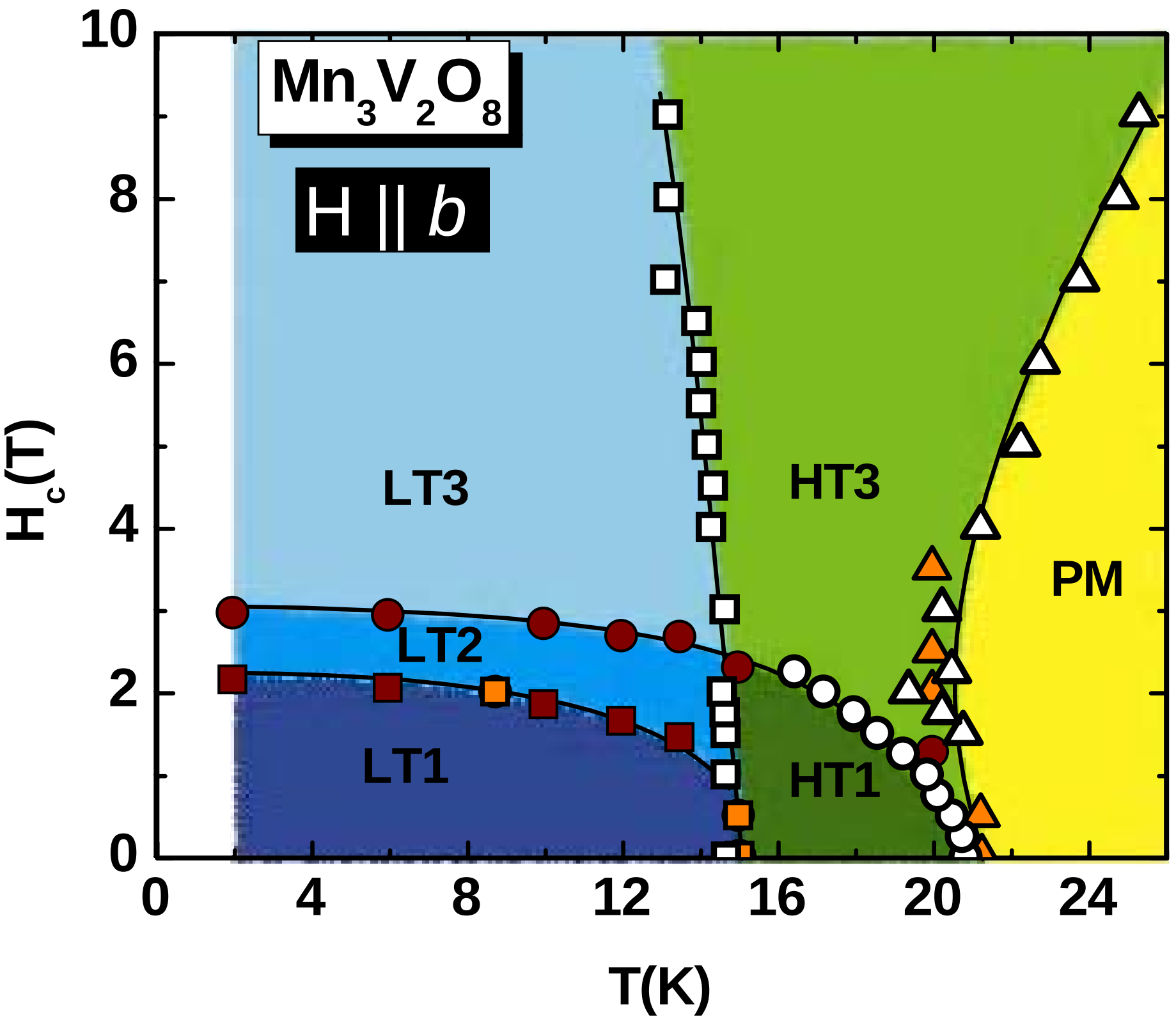